\documentclass[conference]{IEEEtran}
\IEEEoverridecommandlockouts

\usepackage{cite}
\usepackage{amsmath,amssymb,amsfonts}
\usepackage{algorithmic}
\usepackage{graphicx}
\usepackage{textcomp}
\usepackage{xcolor}
\usepackage{subcaption}
\usepackage{tikz}
\usepackage{mwe}
\usepackage{orcidlink}
\usepackage{bibentry}
\usepackage{hyperref}
\usepackage[table]{xcolor}
\usepackage{physics}
\usepackage{relsize}

\newcommand{\subfigimg}[3][,]{%
  \setbox1=\hbox{\includegraphics[#1]{#3}}
  \leavevmode\rlap{\usebox1}
  \rlap{\hspace*{-12pt}\raisebox{\dimexpr\ht1-2\baselineskip + 12pt}{#2}}
  \phantom{\usebox1}
}

\newcommand{\figref}[1]{Fig.~\ref{#1}}

\newcommand{\eqnref}[1]{Equation~\eqref{#1}}
\newcommand{\appref}[1]{Appendix~\ref{#1}}

\def\BibTeX{{\rm B\kern-.05em{\sc i\kern-.025em b}\kern-.08em
T\kern-.1667em\lower.7ex\hbox{E}\kern-.125emX}}
\begin{document}

\title{NEGF Modeling of Impact Ionization in Semiconductor Avalanche Photodiodes for Quantum Networking}

\author{
  \IEEEauthorblockN{Colin Burdine\orcidlink{0000-0001-9699-5804}\IEEEauthorrefmark{1}\IEEEauthorrefmark{2}\IEEEauthorrefmark{3},
    Nischal Binod Gautam \orcidlink{0009-0001-1230-9967} \IEEEauthorrefmark{1}\IEEEauthorrefmark{2}\IEEEauthorrefmark{4},
  Enrique P. Blair\orcidlink{0000-0001-5872-4819}\IEEEauthorrefmark{1}\IEEEauthorrefmark{5}}

  \IEEEauthorblockA{\IEEEauthorrefmark{1}
    \textit{Electrical and Computer Engineering Department},
    \textit{Baylor University}, Waco, TX, USA \\
  }
  \IEEEauthorblockA{\IEEEauthorrefmark{2}
    These authors contributed equally to this work.
  }
  \IEEEauthorblockA{\IEEEauthorrefmark{3}
    Corresponding Author: Colin\_Burdine1@baylor.edu
  }

  \IEEEauthorblockA{\IEEEauthorrefmark{4}
    Nischal\_Gautam1@baylor.edu
  }
  \IEEEauthorblockA{\IEEEauthorrefmark{5}
    Enrique\_Blair@baylor.edu
  }
}

\maketitle

\vspace{-40em}
\begin{abstract}


  We present an atomistic quantum transport simulation framework based on the Non-Equilibrium Green's Function (NEGF) formalism to model impact ionization in semiconductor avalanche devices, with direct relevance to near-term quantum networking applications. Conventional descriptions of avalanche breakdown rely predominantly on semiclassical simulation methods, such as local ionization coefficients, semiclassical carrier trajectories, or Monte Carlo sampling, all of which implicitly assume weak correlations and mean-field electronic interactions. These assumptions break down in nanoscale, high-field junctions where carrier multiplication emerges from strongly non-equilibrium, energy-resolved scattering processes.

  Our approach formulates impact ionization as a multi-particle self-energy within NEGF, enabling a non-perturbative, energy- and atomic orbital-resolved description of carrier multiplication directly from the device spectral function. This formulation captures strongly inelastic scattering processes beyond semiclassical approximations and is implemented in a matrix-based real-space representation suitable for nanoscale device modeling. Using a model semiconductor structure under high electric fields, we demonstrate the emergence of carrier multiplication and analyze its dependence on energy-resolved transport and nonequilibrium charge distributions. The framework provides insight into microscopic mechanisms governing avalanche processes and their impact on device performance.

  Because single-photon detectors form the measurement interface between fragile optical quantum states and classical network operations, a microscopic description of avalanche onset is essential for connecting device physics to quantum-network performance. The results establish a transport baseline for self-consistent calculations of the impact-ionization self-energy and carrier multiplication. By resolving the available and occupied states that underlie avalanche onset, this framework provides a route toward predictive modeling of silicon single-photon avalanche detectors and avalanche photodiodes used in quantum-network receivers.

\end{abstract}

\begin{IEEEkeywords}
  Non-Equilibrium Green's Function, Quantum Kinetic Framework, Quantum Transport, Inelastic Scattering, Electron-Electron Scattering, Impact Ionization, Single-Photon Detection, Avalanche Photodiodes, First-Principles Device Modeling, Carrier Multiplication
\end{IEEEkeywords}

\section{Introduction}

Single photons are the natural flying carriers of quantum information because they travel rapidly through fiber and free space, interact weakly with the environment, and can be prepared, routed, interfered, and measured with mature optical technology \cite{gisin2002quantum, feng2022silicon}. In a quantum network, a photon is a physical qubit whose logical value is encoded in an optical degree of freedom, distributed through a lossy channel, and decoded at the receiver. Polarization encoding maps logical states to horizontal/vertical, diagonal/antidiagonal, or circular polarization bases while path and dual-rail encoding use the presence of one photon in one of two spatial modes. Time-bin encoding uses early and late temporal modes, orbital angular momentum (OAM) and vector-vortex states provide higher-dimensional alphabets, and photon-number encodings use occupation states of one or more optical modes \cite{thiel2024time, wang2022orbital, suprano2023orbital, vajner2025exploring, zhang2025survey, nerenberg2025photon}. Decoding reverses this preparation by using polarizing beam splitters, interferometers, mode sorters, phase settings, and time-resolved detection to choose a measurement basis and convert the arriving quantum state into a classical outcome~\cite{pica2022step}. The detector is therefore not a passive counter at the edge of the experiment. It is part of the measurement operation that turns a distributed optical state into a usable bit, syndrome, herald, or feed-forward signal.

This measurement role is critical in quantum key distribution (QKD), where detection events determine which transmitted quantum states become secret key material. In BB84, correlated detections after basis reconciliation, error estimation, and privacy amplification form the final key \cite{bennett2014quantum, gisin2002quantum, scarani2009security}, while entanglement-based and decoy-state protocols use detector outcomes to reveal eavesdropping and separate single-photon contributions from multiphoton emissions \cite{ekert1991quantum, lo2005decoy}. Detector efficiency, dark counts, timing jitter, dead time, afterpulsing, and saturation therefore enter directly into secret-key rate, maximum distance, and security margins \cite{scarani2009security, grunenfelder2023fast, tsochev2026single}. Measurement-centered protocols make this dependence even clearer: BBM92 and measurement-device-independent QKD rely on detector events and Bell-state measurements to generate correlations while managing detector-side assumptions \cite{bennett1992quantum, lo2012measurement}. Device physics is therefore part of the rate-distance budget, not a detail added after the protocol is designed.

Single-photon detection is central to teleportation and entanglement-based networking, where Bell-state measurements, classical communication, and conditional corrections transfer or extend quantum states \cite{bennett1993teleporting, strobel2024quantum}. In optical implementations, beam splitters, phase stability, photon indistinguishability, and coincidence detection determine heralding fidelity, while missed or spurious clicks create failure or false-success events. Entanglement swapping and repeater protocols such as DLCZ therefore rely on detector quality to extend entanglement beyond direct links and maintain useful end-to-end rates \cite{zukowski1993event, pan1998experimental, briegel1998quantum, duan2001long}.

Photonic quantum computing places similar demands on single-photon generation, distribution, and detection. The KLM protocol showed that scalable computation can be performed with single photons, passive linear optics, auxiliary states, photon counting, and measurement-conditioned feed-forward rather than strong optical nonlinearities \cite{knill2001scheme, kok2007linear}. Later linear-optical and measurement-based schemes use cluster states, fusion measurements, and repeated single-photon measurements to drive computation \cite{raussendorf2003measurement}. In these architectures, detector clicks become computational events: inefficiency produces loss, dark counts produce logical or leakage errors, timing jitter limits interference visibility, and finite photon-number resolution restricts available gates and encodings \cite{kok2007linear, nishio2025multiplexed, kaneda2026active}.

The network-level picture that emerges is a source--channel--receiver chain. Integrated silicon photonics is attractive for this chain because it can combine sources, phase shifters, interferometers, multiplexers, mode converters, and detectors on compact chips compatible with scalable fabrication \cite{feng2022silicon, thiel2024time}. High-dimensional encodings based on OAM, time bins, path, polarization, and photon number can increase channel capacity, support multiplexing, or improve error tolerance, but they also require receivers that preserve mode selectivity while registering sparse events with high efficiency \cite{wang2022orbital, suprano2023orbital, vajner2025exploring, zhang2025survey, nishio2025multiplexed}. In free-space, satellite, and CubeSat links, detector size, weight, power, radiation tolerance, and thermal operating conditions become as important as raw efficiency \cite{tsochev2026single}. For visible and near-visible wavelengths, silicon single-photon avalanche diodes (SPADs) remain especially attractive because they offer mature fabrication, compact integration, high-speed electronic readout, and favorable system-level engineering compared with many cryogenic detector platforms \cite{cusini2022historical, an2025silicon, raimondi20252, liang2026ghz}.

The internal gain mechanism that enables single-photon sensitivity in avalanche photodiodes is \textit{impact ionization}: a carrier accelerated by a strong electric field gains enough kinetic energy to excite an additional electron--hole pair, initiating a multiplication chain. This microscopic process is indispensable for Geiger-mode operation. A single absorbed photon creates a primary carrier, the high-field region converts that carrier into a macroscopic avalanche, and the readout circuit registers a digital pulse. The same stochastic multiplication process, however, also introduces excess noise, breakdown-time variation, afterpulsing, and a dependence of detection probability on where and when the photon is absorbed. These effects propagate upward into QKD error rates, teleportation and swapping fidelities, heralding efficiencies, time-bin discrimination, and the clock rates of photonic computing protocols \cite{grunenfelder2023fast, strobel2024quantum, kok2007linear, petticrew2017avalanche}.

\begin{figure}[htbp]
  \centering
  \includegraphics[width=0.45\textwidth]{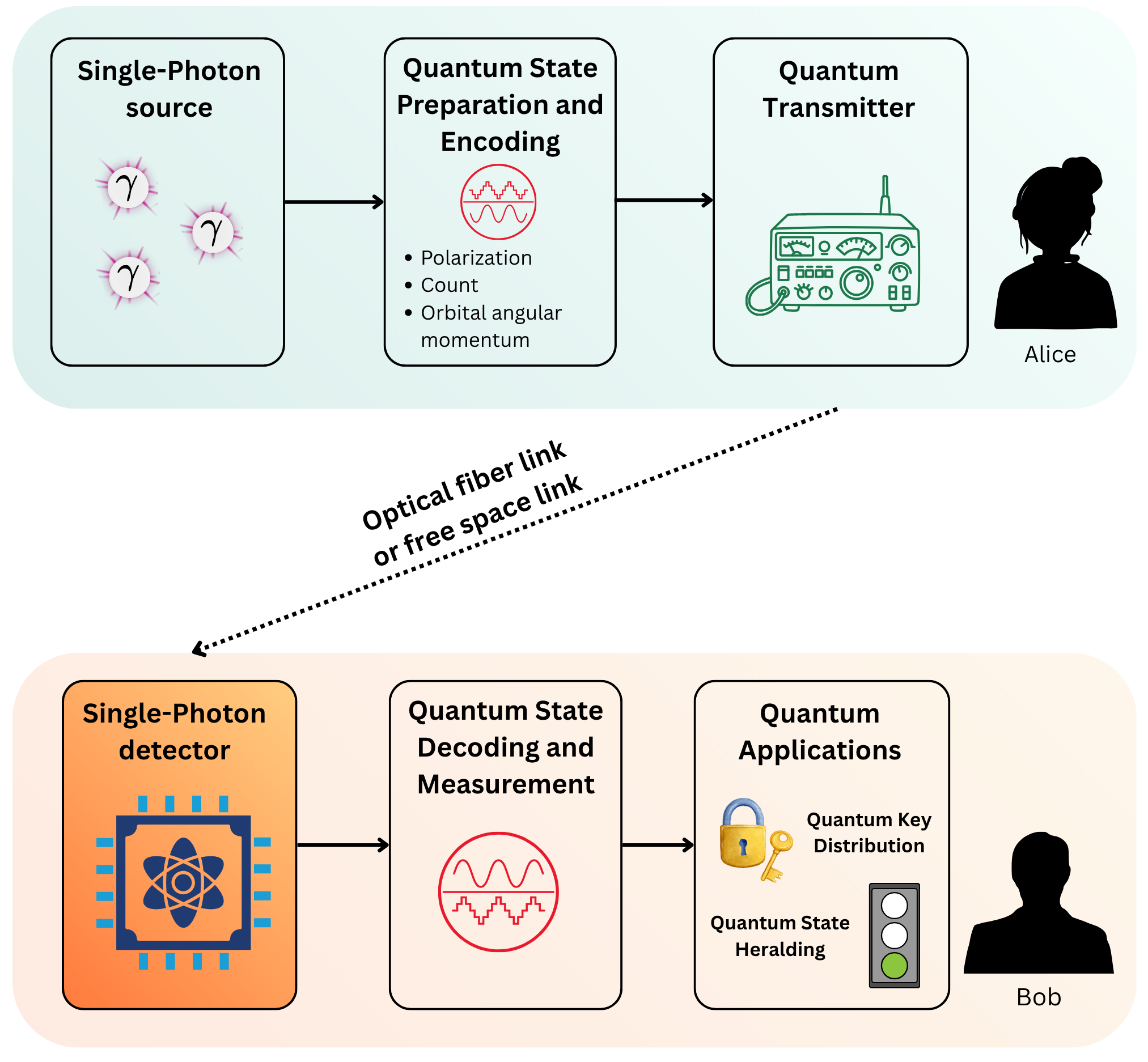}
  \caption{The simplified  block diagram of a representative quantum networking shows the various components involved in it.
  In this work, our focus is on the single-photon detector block, specifically silicon avalanche-based detection.}
  \label{block}
\end{figure}

Historically, APD and SPAD simulations have used drift-diffusion models, local ionization coefficients, Monte Carlo transport, and phenomenological dead-space descriptions \cite{qian2023modeling, dollfus2022avalanche}. While effective at larger scales, these approaches become less adequate in nanometer-scale multiplication regions, where atomistic band structure, tunneling, state broadening, and nonequilibrium occupation shape avalanche onset. Because the detector block in Fig.~\ref{block} converts fragile optical information into a timed electrical event, silicon SPAD modeling must capture high-field transport through the junction. We therefore use a generic silicon p--n or p--i--n-like structure as a transparent but physically relevant starting point for studying depletion-region transport, contact injection, and avalanche-driving fields.

The non-equilibrium Green's function formalism is uniquely suited to the sub-nm device regime because it describes transport in open quantum systems through propagators and correlation functions rather than semiclassical trajectories. Contact injection, coherent propagation, state broadening, and inelastic scattering can all be expressed in a unified framework \cite{datta2018lessons}. Recent matrix quantum-kinetic work has further shown that impact ionization itself can be written as a multi-Green's function self-energy, allowing carrier multiplication to be computed from the underlying nonequilibrium spectral structure of the device instead of inserted phenomenologically \cite{ahmed2025matrix}. For quantum networking detectors, that possibility is appealing because it offers a route from device geometry and material choice to important properties such as gain onset, carrier multiplication, and timing statistics.

This paper develops a NEGF-based modeling workflow for a silicon avalanche structure intended as a physically transparent starting point for single-photon detector simulation. The contributions of this paper are threefold. First, it places the detector problem in the context of quantum networking, making clear why microscopic APD models matter at the system level. Second, it formulates a silicon p--i--n-like avalanche device in a matrix NEGF formalism that separates contact injection, electrostatics, and ionization scattering into interpretable pieces. Third, it uses the simulated figures in this work to identify the spectral and spatial conditions required for impact ionization to become efficient. Together, these results lay the foundation for a predictive, detector-oriented quantum transport model of silicon SPADs.


\section{Background}

\begin{figure*}[htbp]
  \centering
  \includegraphics[width=\textwidth]{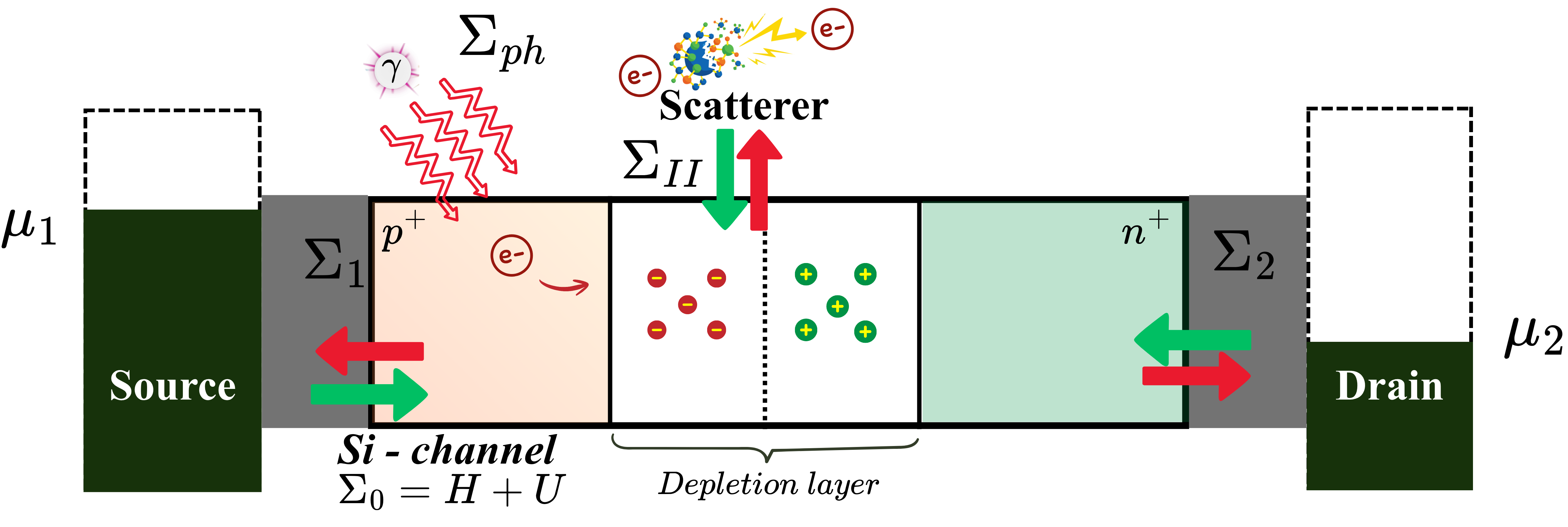}
  \caption{The quantum transport model of the generic silicon p--n junction device enables the modeling of incoherent NEGF transport using the channel Hamiltonian matrix $H$, the self-energy matrices $\Sigma_1$ and $\Sigma_2$ that describe coupling to the contacts, $\Sigma_{II}$ that describes the impact ionization scattering self-energy, the self-energy term for photon scattering $\Sigma_{ph}$, and $\Sigma_0$ that accounts for interactions within the channel. The channel is connected to source and drain contacts and at each terminal the influx is described by the electron in-scattering function $\Sigma^{\text{in}}$, which injects contact electrons into hole states $G^>$ in the channel, while the outflow is governed by the out-scattering function $\Sigma^{\text{out}}$, which removes electrons from channel states $G^<$. The contact terminals follow Fermi--Dirac distributions determined by their respective electrochemical (quasi-Fermi) energies dictated by $\mu_1$ and $\mu_2$.  }
  \label{fig:apd_device}
\end{figure*}

\subsection{Non-Equilibrium Green's Functions}

This section summarizes the energy-resolved formulation of non-equilibrium Green's function (NEGF) theory used throughout this work. A detailed derivation from the time-domain many-body formalism is provided in \appref{app:negf_review}; here, we restrict attention to the steady-state limit and introduce only the quantities required for atomistic transport modeling.

We begin by specifying the second-quantized Hamiltonian,
\begin{equation}
  \hat{H} =
  \underbrace{\sum_{ij} h_{ij}\hat{c}_i^\dagger \hat{c}_j}_{\hat{H}_0}
  +
  \underbrace{\sum_{ijkl}\frac{1}{2}V_{ijkl}
  \hat{c}_i^\dagger \hat{c}_j^\dagger \hat{c}_k \hat{c}_l}_{\hat{H}_{int}} .
  \label{eqn:Hamiltonian}
\end{equation}
Above, the operators $\hat{c}_i^\dagger$ and $\hat{c}_i$ create and annihilate an electron in mode $i$, and the matrix elements $h_{ij}$ define the single particle dynamics, including the band structure and electrostatic potential. The coefficients $V_{ijkl}$ describe two body interactions, including Coulomb and scattering between electronic states. In the NEGF framework, the effects of the interaction terms $V_{ijkl}$ and open boundaries are incorporated through self-energy operators, allowing the problem to be recast in terms of single-particle Green's functions.

Under steady-state conditions, all two-time Green's functions are time-translation invariant, and thus admit a Fourier representation in energy. The central quantities are the retarded, advanced, and lesser Green's functions, denoted $G^R(E)$, $G^A(E)$, and $G^{<}(E)$, respectively. Since the retarded and advanced Green's functions are Hermitian adjoints in this limit (i.e., $G^{R} = (G^{A})^\dagger$), we use the notation
\begin{equation}
  G(E) \equiv G^{R}(E) \quad \text{and}\quad G^\dagger(E) \equiv G^{A}(E)
\end{equation}
for brevity. In the energy domain, the retarded Green's function $G(E)$ satisfies the Dyson equation
\begin{equation}
  G(E) = [(E + i\eta)I - H - \Sigma(E)]^{-1}
  \label{eqn:dyson}
\end{equation}
where $H$ is the matrix representation of $h_{ij}$, $\eta = 0^+$ enforces the retarded boundary condition (i.e., causality), and $\Sigma(E)$ is the retarded self-energy, which incorporates both interaction and boundary effects. As with $G/G^\dagger$, we adopt the shorthand
\begin{equation}
  \Sigma(E) \equiv \Sigma^R(E) \quad \text{and}\quad
  \Sigma^\dagger(E) \equiv \Sigma^A(E)
\end{equation}
to denote the retarded and advanced self-energies, respectively. The carrier occupation statistics of a system of interest are captured in $G^{<}(E)$, whereas particle injection, ejection, and scattering are described in $\Sigma^{<}(E)$. These two operators are related via the Keldysh relations:
\begin{align}
  G^{<}(E) &= G(E)\Sigma^{<}(E)G^\dagger(E)   \label{eqn:keldysh1} \\
  G^{>}(E) &= G(E)\Sigma^{>}(E)G^\dagger(E)
  \label{eqn:keldysh2}
\end{align}
The Green's function $G(E)$ itself encodes the energy-resolved density of available carrier states. This is represented in the system's spectral function, given by
\begin{equation}
  A(E) = i(G(E) - G^\dagger(E)).
  \label{eqn:spectral_fn}
\end{equation}

Altogether, these relations define a compact, closed algebraic system in the energy domain, once the self-energies $\Sigma(E), \Sigma^<(E)$ are specified. This provides a useful foundation for modeling quantum transport in open systems.


\subsection{Energy-Resolved NEGF for Open Semiconductor Devices}

In the NEGF framework, quantum transport in semiconductor devices is typically formulated by partitioning the system into a finite active region (e.g., the intrinsic region of a silicon p--i--n junction) coupled to semi-infinite contacts, such as the device shown in \figref{fig:apd_device}. The open-boundary nature of the problem is incorporated through self-energies $\Sigma$, which account for both carrier exchange with contacts and internal scattering processes. This perspective is essential for reverse-biased junctions, where nonequilibrium transport and carrier multiplication are driven by contact injection and subsequent scattering within the depletion region.

In a semiconductor device, the Hamiltonian $H$ includes both the underlying band structure and an electrostatic potential profile $U(x)$. In the case of silicon APDs, which usually consist of strongly doped p--i--n devices (or similar structures) under strong reverse bias, $U(x)$ plays a crucial role in representing the depletion-region potential that bends the conduction and valence bands and establishes the internal electric field.

In an APD, the total self-energy is decomposed as
\begin{equation}
  \Sigma(E) = \Sigma_L(E) + \Sigma_R(E) + \Sigma_{\mathrm{scatt}}(E),
\end{equation}
with an analogous decomposition for $\Sigma^\dagger(E)$. The contact self-energies $\Sigma_{L,R}$ describe carrier escape into, and injection from, the reservoirs, while $\Sigma_{\mathrm{scatt}}$ captures internal processes such as phonon scattering and impact ionization.

The contact couplings that inject and eject carriers are characterized by the broadening matrices
\begin{equation}
  \Gamma_\alpha(E) = i\left(\Sigma_\alpha(E) - \Sigma_\alpha^\dagger(E)\right),
  \quad \alpha \in \{L,R\},
\end{equation}
which determine the rate at which states in the device hybridize with the contacts. Carrier injection is encoded through the lesser self-energy,
\begin{equation}
  \Sigma^{<}(E) = \sum_{\alpha \in \{L,R\}} i f_\alpha(E)\,\Gamma_\alpha(E) + \Sigma^{<}_{\mathrm{scatt}}(E),
\end{equation}
where $f_\alpha(E)$ are Fermi--Dirac distributions set by the contact electrochemical potentials. Under applied bias, the difference between \( f_L \) and \( f_R \) drives a nonequilibrium carrier population across the device.

The resulting carrier distributions encoded in $G^{<,>}$ are obtained from \eqnref{eqn:keldysh1},
which distinguishes between available states and their occupation. Similarly, \eqnref{eqn:spectral_fn}
describes the density of available states, while $G^{<}(E)$ encodes which of those states are populated. In a reverse-biased silicon junction, this distinction is critical, since avalanche processes require both accessible high-energy states (captured by $A(E)$) and sufficient nonequilibrium occupation (captured by $G^{<}(E)$).

Macroscopic device observables can also be derived from $\Sigma^{<}$ and $G^{<}$. For example, the local carrier density is given by
\begin{equation}
  n(\mathbf{r}) = -\frac{i}{2\pi} \int dE\, G^{<}(\mathbf{r}, \mathbf{r}; E).
\end{equation}

In the limit of coherent carrier propagation (e.g., when scattering is minimal, and the mean free path is large), transport reduces to the Landauer--B\"uttiker form \cite{buttiker1985generalized, ahmed2025matrix}, from which the device's transmission spectrum $T(E)$ can be derived:
\begin{equation}
  T(E) = \mathrm{Tr}\left[\Gamma_L(E)\,G(E)\,\Gamma_R(E)\,G^\dagger(E)\right],
  \label{eqn:transmission}
\end{equation}
yielding the current
\begin{equation}
  I = \frac{q}{\hbar} \int \frac{dE}{2\pi}\, T(E)\left[f_L(E) - f_R(E)\right].
\end{equation}
However, in the presence of scattering, the current is dependent on the combined occupations of all leads, as well as any scattering sources present. In this case, the terminal current $I_\alpha$ is given by the more general Meir-Wingreen formula \cite{meir1992landauer, ghosh2016nanoelectronics, ahmed2025matrix}:
\begin{equation}
  I_\alpha =
  \frac{q}{\hbar}
  \int \frac{dE}{2\pi}\,
  \mathrm{Tr}\!\left[
    \Sigma_\alpha^{<}(E)G^{>}(E)
    -
    \Sigma_\alpha^{>}(E)G^{<}(E)
  \right].
\end{equation}

This formulation provides a direct connection between the device Hamiltonian (including band bending), the open-boundary conditions imposed by contacts, and the resulting nonequilibrium carrier dynamics. The inclusion of \( \Sigma_{\mathrm{scatt}} \) enables the treatment of dissipative and multiplication processes within the same energy-resolved framework, forming the basis for modeling avalanche behavior in semiconductor devices.

\subsection{Impact Ionization Self-Energy}

Avalanche transport requires a scattering mechanism capable of converting a single high-energy carrier into multiple carriers. Within the NEGF framework, we incorporate this process through an impact-ionization self-energy, which captures correlated two-particle scattering beyond simple reservoir injection. Unlike contact self-energies, which depend on equilibrium distributions, the impact-ionization term depends explicitly on products of band-restricted lesser and greater Green's functions, reflecting the underlying four-state collision process in the interaction Hamiltonian $H_{int}$. In compact form, the lesser self-energy for impact ionization takes the form \cite{ahmed2025matrix}:
\begin{align}
  \Sigma_{II}^{>} &= D_0 \odot \left(\sum_{b,b'} [G_b^{min} \ast G_b^{maj} G_{b'}^{maj}]\right) \label{eqn:sigma_ii_g}\\
  \Sigma_{II}^{<} &=  D_0 \odot \left(\sum_{b,b'} [G_{b'}^{maj} \ast G_b^{min} G_{b}^{min}]\right) \label{eqn:sigma_ii_l}\\
  b, b' \in \{c,v\} \notag
\end{align}
where $D_0$ is an effective interaction constant, $\odot$ denotes element-wise multiplication, and the convolution is defined by
\begin{equation}
  \begin{aligned}
    &[A \ast BC] (E) = \iiint \frac{dE'}{2\pi}\frac{dE''}{2\pi}\frac{dE'''}{2\pi} \Big[  \\
    & \qquad A(E')B(E'')C(E''') \delta(E + E' - E'' - E''') \Big]
  \end{aligned}
\end{equation}
In equations \eqref{eqn:sigma_ii_g} and \eqref{eqn:sigma_ii_l}, $G_b^{\mathrm{maj}}$ and $G_b^{\mathrm{min}}$ denote the band-resolved Green's function components weighted by occupied (majority) and unoccupied (minority) spectral contributions, respectively, thereby separating the availability of carriers from the availability of final states within the conduction or valence band $b \in \{c,v\}$.

\begin{figure}[htbp]
  \centering
  \includegraphics[width=0.45\textwidth]{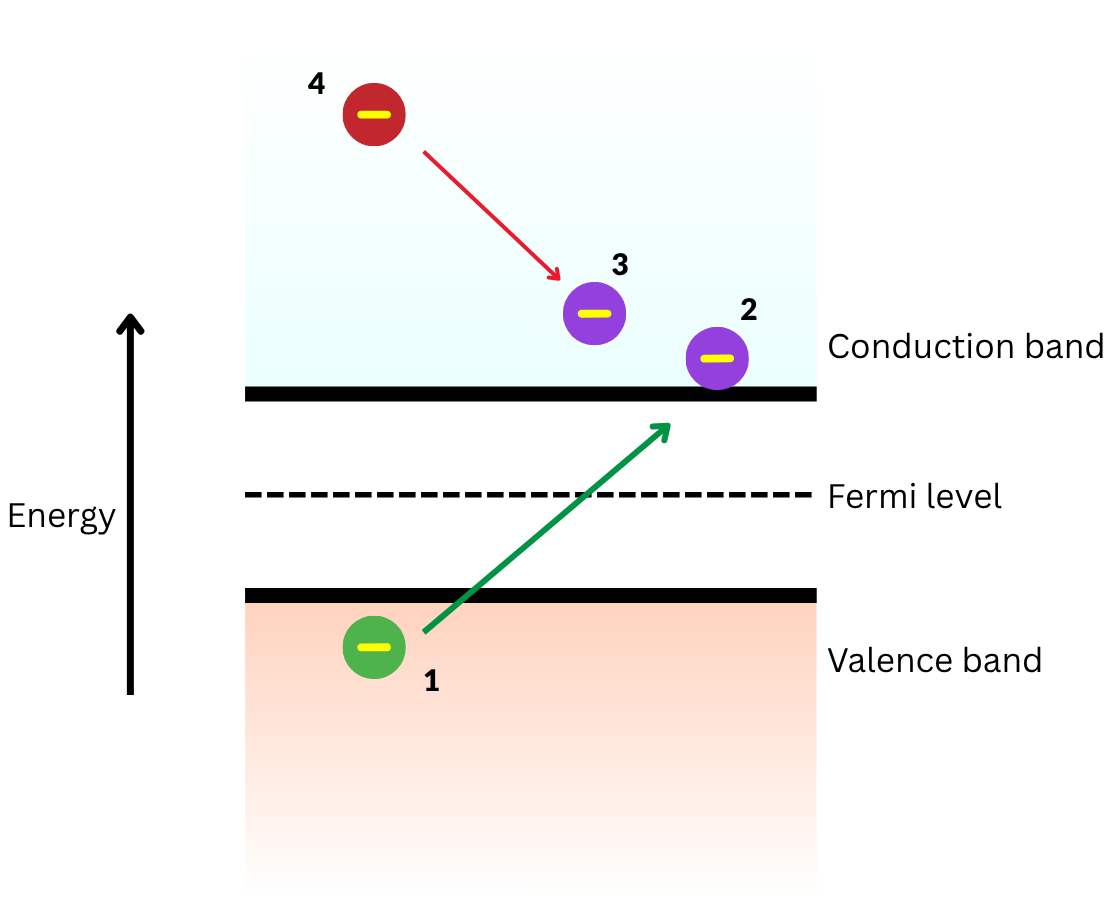}
  \caption{A very high energy electron (4) collides with a top of the valence-band electron (1), scattering both into empty states (2, 3), and resulting in two lower-energy conduction electrons near the band bottom along with holes at the original states 1 and 4, giving us the four energy levels. The energy levels satisfy energy conservation, $E_4 - E_3 = E_2 - E_1$. }
  \label{fig:four_level_energy}
\end{figure}

This structure enforces energy conservation while coupling occupied and unoccupied states across bands, as shown in \figref{fig:four_level_energy}. Physically, an ionization event requires: (i) an energetic carrier, (ii) an available partner state, and (iii) accessible final states consistent with conservation laws. The resulting self-energy therefore depends on the joint spectral and occupation overlap encoded in the Green's functions, rather than on local field strength alone.

In a strongly reverse-biased silicon p--i--n junction, this distinction is critical. Band bending across the depletion region creates a spatially varying spectral landscape such that the probability of impact ionization is governed by the local density of states and nonequilibrium carrier populations.

\subsection{The Self-consistent Born Approximation}

As discussed above, scattering in the NEGF formalism is incorporated through energy-dependent self-energy contributions to the device Green's functions. The self-consistent Born approximation (SCBA) closes the problem by expressing the scattering correlation self-energies as functionals of the unknown correlation Green's functions, allowing for an iterative fixed-point method to be applied \cite{williams2009exploring}.

In the SCBA method, each scattering mechanism provides a map $(G^<,G^>)\mapsto(\Sigma_{\mathrm{scatt}}^<,\Sigma_{\mathrm{scatt}}^>)$; this is implemented by summing contributions from a list of scatterers,
\begin{equation}
  \Sigma_{\mathrm{scatt}}^{<,>}(E)=\sum_s \Sigma_{s}^{<,>}\!\big[G^<(E),G^>(E)\big].
\end{equation}

A key detail is how the retarded scattering self-energy $\Sigma_{\mathrm{scatt}}$ is constructed from the correlation (broadening) content in a way that preserves causality and spectral weight. Defining the scattering broadening
\begin{equation}
  \Gamma_{\mathrm{scatt}}(E) = i[\Sigma_{\mathrm{scatt}}^>(E)-\Sigma_{\mathrm{scatt}}^<(E)],
\end{equation}
we apply the Kramers--Kronig relations to tie together the imaginary and real parts of $\Sigma_{\mathrm{scatt}}$:
\begin{equation}
  \Sigma_{\mathrm{scatt}}(E)= -\frac{i}{2}\Gamma_{\mathrm{scatt}}(E) + \mathcal{H}\!\left[\Gamma_{\mathrm{scatt}}\right](E),
\end{equation}
where $\mathcal{H}[f]$ is the Hilbert transform
\begin{equation}
  \mathcal{H}[f](E)=\frac{1}{\pi}\,\mathcal{P}\!\!\int_{-\infty}^{\infty}\frac{f(E')}{E-E'}\,dE'.
\end{equation}
This Kramers--Kronig completion enforces retarded causality, ensuring that dispersive energy shifts (real part) are consistent with the dissipative broadening (imaginary part). In turn, the resulting retarded Green's function produces a physically consistent spectral function \eqref{eqn:spectral_fn}.

To obtain a self-consistent set of $G, G^{<}, G^{>}, \Sigma, \Sigma^{<}$, and $\Sigma^{>}$, an SCBA solver performs a fixed-point iteration over $\Sigma_{\mathrm{scatt}}^{<,>}$. Starting from an initial guess, it computes $\Gamma_{\mathrm{scatt}}$, forms $\Sigma_{\mathrm{scatt}}^R$ using the Hilbert transform recomputes $G^R$, updates $G^{<,>}$ via the Keldysh relation, and then evaluates new $\Sigma_{\mathrm{scatt,new}}^{<,>}$ from the scatterers. To stabilize convergence it applies linear mixing,
\begin{equation}
  \Sigma_{\mathrm{scatt}}^{<,>}\leftarrow (1-\alpha)\,\Sigma_{\mathrm{scatt}}^{<,>}+\alpha\,\Sigma_{\mathrm{scatt,new}}^{<,>},
\end{equation}
with an optional adaptive decrease of $\alpha$ if a normalized SCBA residue exhibits oscillatory behavior.

\section{Results and Discussion}

\begin{figure}[h]
  \centering
  \begin{tabular}{c c}
    \subfigimg[width=0.8\linewidth, trim={0.2in 0 0 0.4in}]{(a)}{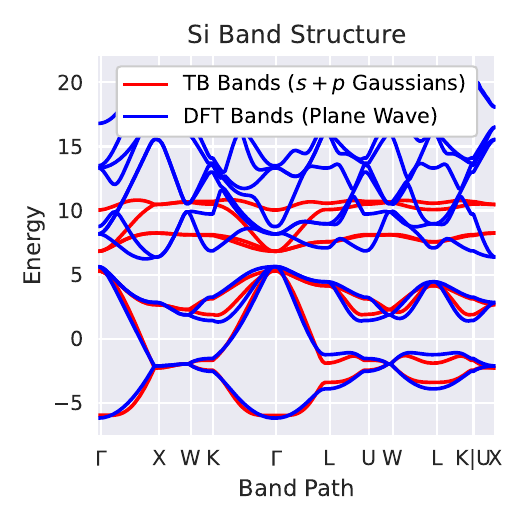} \\
    \subfigimg[width=0.8\linewidth, trim={0.2in 0 0 0.4in}]{(b)}{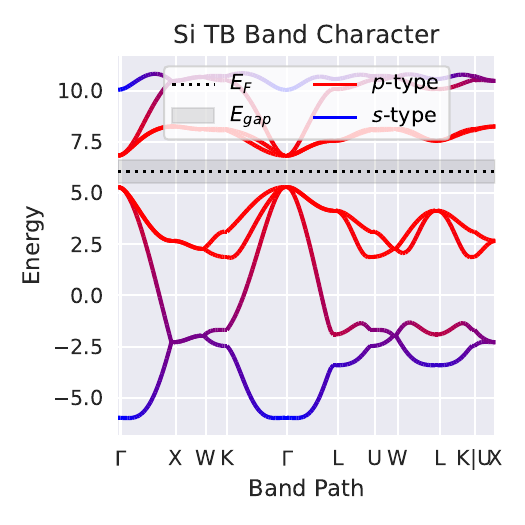}
  \end{tabular}
  \caption{(a) The blue curves in the silicon band structure along high-symmetry points in the Brillouin zone show the reference DFT plane-wave bands, while the red curves show the fitted equivariant tight-binding bands used in the device Hamiltonian. The comparison shows how the reduced tight-binding model captures the low-energy band features needed for transport while remaining compact enough for NEGF calculations. (b) In the orbital character of the silicon tight-binding band structure, red and blue curves show dominant $p$- and $s$-orbital character, where conduction/valence band states correspond to $sp^3$-bonding/antibonding states.}
  \label{fig:si_bands}
\end{figure}

Now that we have introduced the relevant NEGF theory, we present results applying this theory in an atomistic computational framework to model impact ionization in nanometer-scale silicon devices. Here, we present preliminary results obtained through a computational implementation. This section presents the atomistic and NEGF quantities used to build the silicon avalanche device model. The results first validate the tight binding representation of silicon, then construct the device geometry and transverse momentum space, and finally show how contact coupling and electrostatics determine the available and occupied states in the channel. These quantities form the transport baseline needed before a fully self consistent impact ionization calculation can be interpreted.

Figure~\ref{fig:si_bands} shows the electronic structure model used in the transport calculation. The fitted tight binding bands reproduce the main low energy structure of the DFT reference bands near the band gap, while using a compact localized basis suitable for matrix NEGF calculations. The higher conduction bands are compressed relative to the DFT reference because the model is optimized for the energy range most relevant to device transport. The orbital character plot shows that the valence and conduction bands near the gap are mainly $p$ like, suggesting that avalanche current and scattering mostly occurs within the $\pi$-bonds between adjacent Si sites. This separation is useful when interpreting carrier injection and impact ionization because the scattering kernel depends on which occupied and empty states are available near the band edges.

\subsection{Device Geometry and Electronic Structure}

\begin{figure}[htbp]
  \centering

  \subfigimg[height=1.4in, trim={1.5in 0.45in 1in 0.75in}, clip]{(a)}{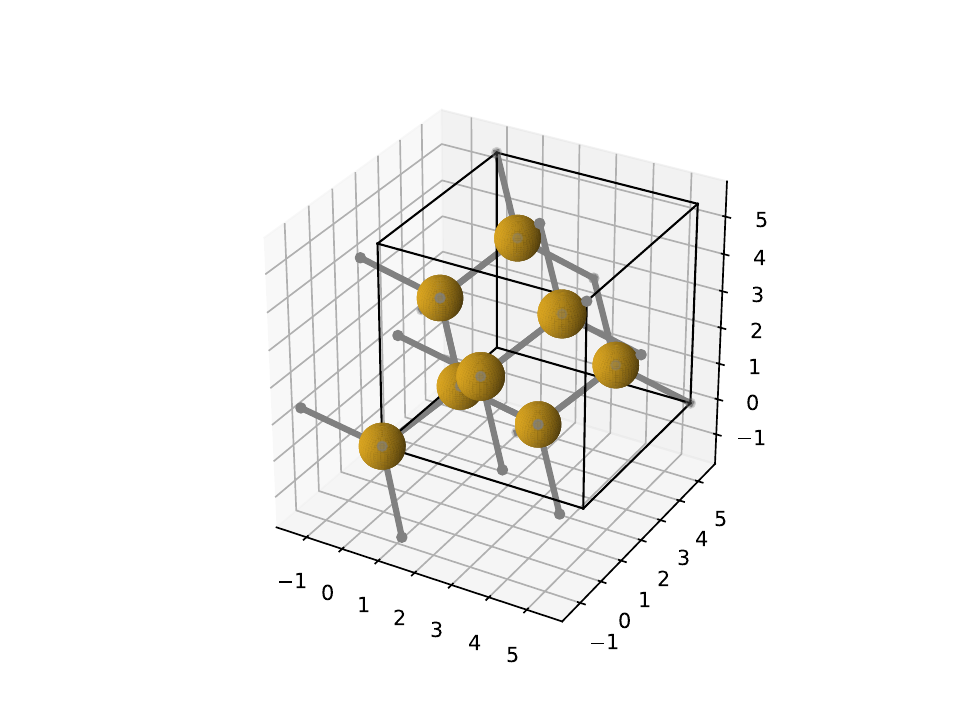} \quad \subfigimg[height=1.4in]{(b)}{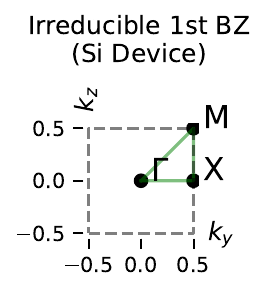} \\
  \vspace{1mm}

  \subfigimg[width=0.9\linewidth, trim={0.25in 0 0.25in 0}]{(c)}{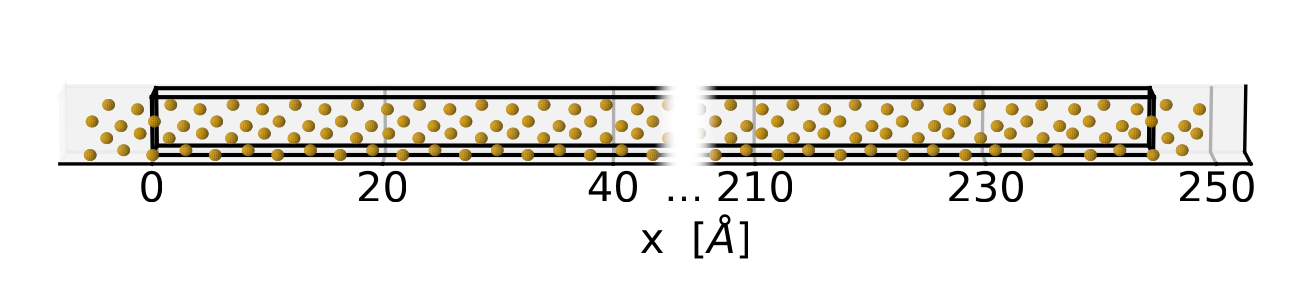}\\
  \caption{(a) Cubic unit cell of silicon containing 8 atoms in a diamond structure. (b) Irreducible 1st Brillouin zone (IBZ), capturing the $\vec{k} = (k_y, k_x)$ momentum degrees of freedom of electrons propagating orthogonal to the device axis. (c) Visualization of the device unit cell of a 25-nm depletion region in a Si APD device, constructed with 45 silicon cells and $p^+$- and $n^+$-doped semi-infinite leads. Periodicity orthogonal to the device axis was acccounted for by sampling the Brillouin zone in (b) with a uniform mesh.}
  \label{fig:device_overview}
\end{figure}

Figure~\ref{fig:device_overview} defines the real space and reciprocal space structure used in the simulation. The silicon unit cell supplies the atomistic basis for the Hamiltonian, and the long device region is built by repeating this cell along the transport coordinate $x$. Periodicity is retained in the directions transverse to transport, so the transverse momenta are sampled over the irreducible Brillouin zone shown in Fig.~\ref{fig:device_overview}(b). This construction keeps the model atomistic along the avalanche direction while accounting for the transverse modes that contribute to current and density of states.

\subsection{Computational Details}

Electronic transport calculations were performed within the non-equilibrium Green’s function (NEGF) formalism using a custom Python-based implementation, with numerics formulated in a tensorized representation and executed via GPU-accelerated linear algebra routines that facilitated solving for the steady-state solutions obtained under the self-consistent Born approximation (SCBA) \cite{haug2008quantum, lake1997single}. Retarded Green’s functions were computed via direct inversion of the Dyson equation, while correlation self-energies were iteratively updated from a modular set of scattering kernels (incorporating impact ionization contributions) until self-consistency was achieved through fixed-point iteration with linear mixing. Surface Green's functions for semi-infinite leads were computed using the iterative Sancho-Rubio decimation scheme \cite{sancho1985highly, sadasivam2017phonon, he2014non}, in which renormalized onsite and coupling matrices are recursively updated until convergence of the surface self-energy is reached.

\begin{table}[h]
  \centering
  \caption{Table of NEGF device and simulation parameters used in this study}
  \rowcolors{2}{gray!20}{white}
  \begin{tabular}{c c c}
    \hline\hline
    Parameter & Value(s) & Description \\
    \hline
    $n_{p^+}$ & $2.5 \cdot 10^{18}$ cm$^{-3}$ & $p^+$ doping \\
    $n_{n^+}$ & $4.0 \cdot 10^{18}$ cm$^{-3}$ & $n^+$ doping \\
    $L$ & 250 \AA & Depletion layer size \\
    $D_0$ & $10$ & Impact ionization coefficient \\
    $\eta$ & $10^{-4}$ & Surface Green's function broadening \\
    $\alpha$ & $0.1$ & SCBA mixing $\alpha$ parameter \\
    \hline\hline
  \end{tabular}
  \label{tab:negf_parameters}
\end{table}

A summarized list of the simulation parameters is presented in Table \ref{tab:negf_parameters}. All simulations were implemented in a fully tensorized manner using PyTorch \cite{paszke2019pytorch}, leveraging GPU acceleration via the CUDA backend for efficient batched linear algebra and Fourier-domain convolutions (e.g., in impact ionization self-energy evaluation). The Sancho-Rubio solver employed adaptive broadening, initializing with an elevated imaginary part ($\eta \sim 10^{-4}$ eV) and progressively reducing it upon intermediate convergence to improve numerical stability. Convergence criteria were defined by a maximum total Frobenius-norm residual below $10^{-4}$ (eV)$^2$, with safeguards against divergence or non-physical values (e.g., NaNs), and a maximum iteration cap of $10^4$. Production simulations and performance benchmarking were carried out on an NVIDIA H200 NVL GPU.

\begin{figure}[htbp]
  \centering

  \begin{tabular}{c c}
    \subfigimg[width=0.7\linewidth, trim={0.2in 0 0 0.4in}]{(a)}{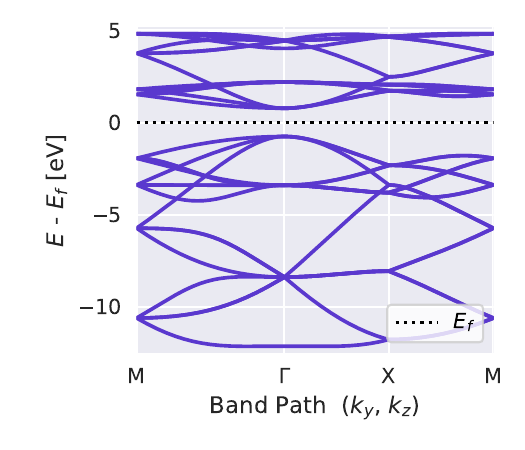}\qquad~ \\
    \subfigimg[width=0.8\linewidth, trim={0.2in 0 0 0.4in}]{(b)}{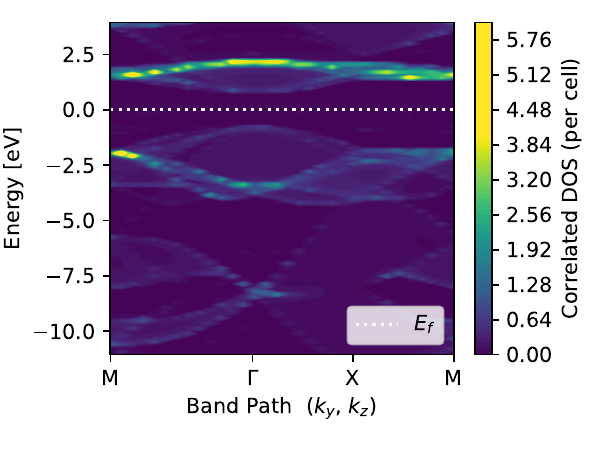}
  \end{tabular}
  \caption{(a): Unit cell band structure of a 2D slab of intrinsic silicon, computed via a localized $s$+$p$-orbital tight-binding model fit to the band structure of bulk silicon. (b): Computed $k$-resolved density of states of a semi-infinite silicon slab, illustrating the effect of band broadening in the NEGF formalism.}
  \label{fig:device_bands}
\end{figure}

\subsection{NEGF Simulation}

Figure~\ref{fig:device_bands} compares the isolated slab band structure with the semi-infinite contact-broadened spectrum. In the isolated case, the allowed states appear as sharp energy bands along the transverse momentum path. After coupling the channel to the lead self-energies, those bands become finite-width spectral features. This broadening is a key NEGF result because an open device does not have perfectly discrete states. Carriers injected from the contacts have finite lifetimes in the channel, and the spectral function records that lifetime through broadened energy support. The strongest spectral weight appears near the band edges, which are the states most relevant for injection, current flow, and the first stages of carrier multiplication.

\begin{figure}[h]
  \centering
  \subfigimg[width=0.9\linewidth, trim={0.25in 0 0.25in 0}]{(a)}{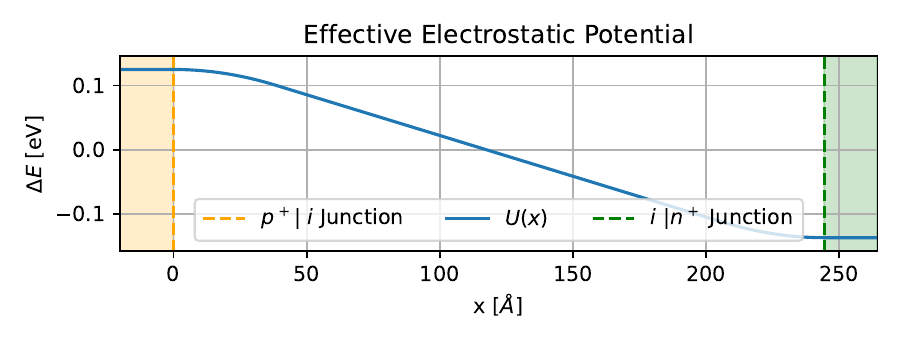}\\[-1mm]
  \subfigimg[width=0.9\linewidth, trim={0 0 0.25in 0}]{(b)}{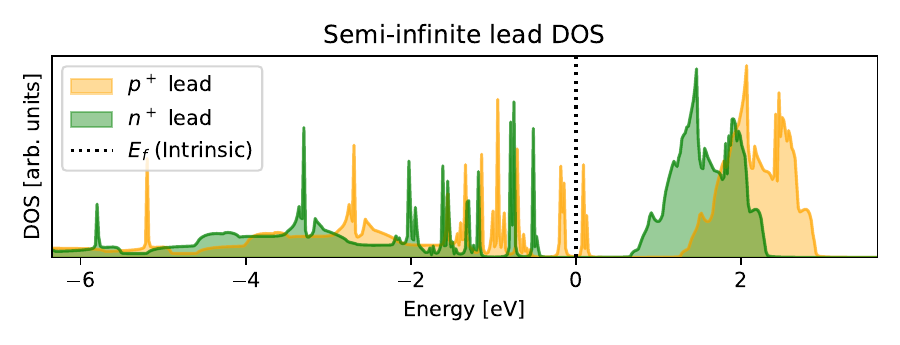}\\[-1mm]
  \subfigimg[width=0.9\linewidth, trim={0 0 0.25in 0}]{(c)}{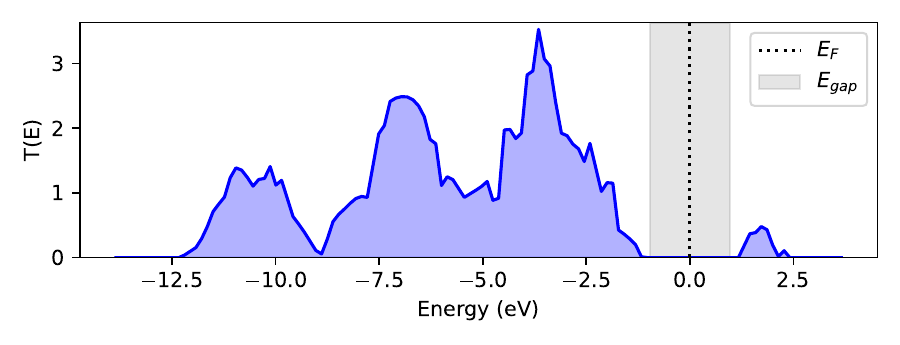}

  \caption{(a): Effective built-in potential of the nm-scale device shown in Figure~\ref{fig:device_overview}, assuming a uniform electrostatic gradient between the $p$-type and $n$-type leads. (b): Density of states computed for the semi-infinite p-type and n-type leads under a small reverse bias ($V_{bias} = -1$  V) to show peak separation. (c): Transmission spectrum of a unit cell of the unbiased device ($V_{bias} = 0$), estimated with \eqnref{eqn:transmission}.}
  \label{fig:electrostatic_transport}
\end{figure}

After establishing the contact-broadened electronic structure, the next step is to include the electrostatic band bending produced by the doped junction. This electrostatic profile determines how the band edges shift across the device and sets the internal field that accelerates carriers through the intrinsic region. The built-in potential is typically estimated from the acceptor $N_A$ and donor $N_D$ concentrations as
\begin{equation}
  V_{bi} = \frac{k_B T}{q} \log\left(\frac{N_A N_D}{n_i^2}\right).
  \label{eqn:builtin_potential}
\end{equation}
The doped leads were then shifted relative to the intrinsic region according to
\begin{align}
  \Delta U_{(p^+)} &= k_BT\log(N_{p^+}/n_i) \\
  \Delta U_{(n^+)} &= - k_BT\log(N_{n^+}/n_i)
  \label{eqn:lead_potentials}
\end{align}
so that the $p^+$ and $n^+$ leads impose the correct relative alignment with the intrinsic channel. These offsets produce the electrostatic landscape shown in Fig.~\ref{fig:electrostatic_transport}(a). This built-in potential bends the bands across the intrinsic region and establishes the internal field that accelerates carriers. The lead density of states in Fig.~\ref{fig:electrostatic_transport}(b) confirms that the doped contacts provide occupied and empty states on opposite sides of the intrinsic Fermi level. This contact asymmetry is what drives injection once bias is applied. The transmission in Fig.~\ref{fig:electrostatic_transport}(c) vanishes inside the band gap, and reappears in the conduction and valence bands where propagating states are available. This behavior is expected for a coherent baseline calculation: the gap suppresses direct transport near $E_F$, while the available lead and channel states away from the gap allow finite transmission.

\begin{figure}
  \centering
  \begin{tabular}{c c}
    \subfigimg[width=0.75\linewidth, trim={0.25in 0.25in 0 0.4in}]{(a)}{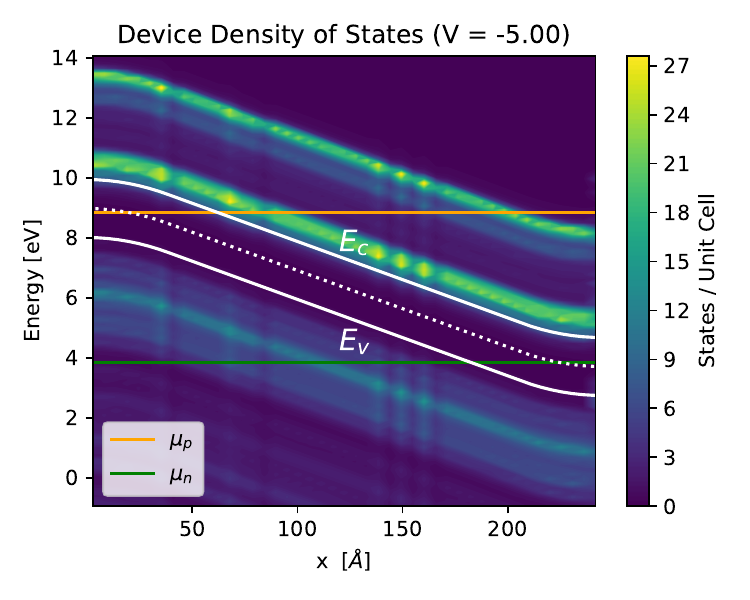} \\
    \subfigimg[width=0.775\linewidth, trim={0.25in 0 0 0.4in}]{(b)}{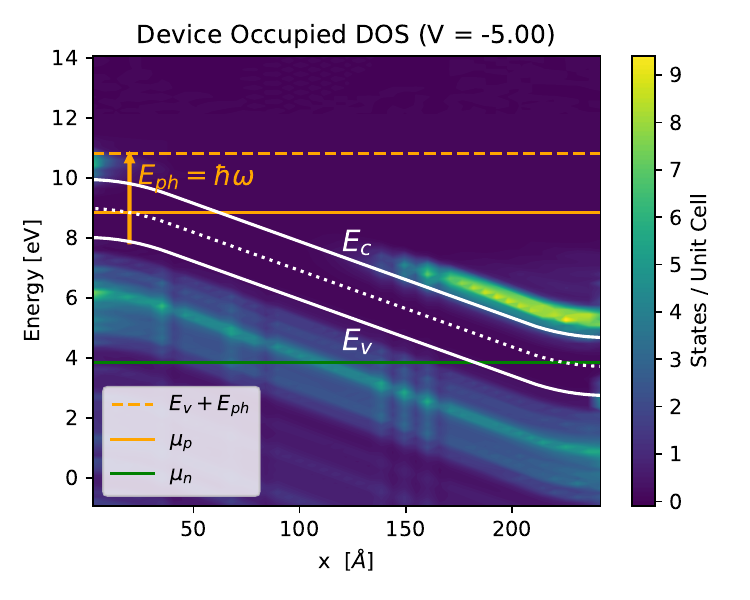}
  \end{tabular}
  \caption{(a) The local density of states (LDOS) shows the available channel states, with the conduction- and valence-band features tilted across the device by the built-in and applied electrostatic field which is necessary to show that carriers can be then accelerated through the high field region. (b) The occupied density of states, computed from the electron correlation function $G^<$, shows the subset of available states populated by contact injection. Their role is to demonstrate that the NEGF simulation has built a physically consistent open-device picture before claiming avalanche behavior. Impact ionization requires both occupied energetic carriers and available final states for scattering.}

  \label{fig:device_ldos_occ}
\end{figure}

Figure~\ref{fig:device_ldos_occ} shows the position- and energy-resolved spectral response of the reverse-biased silicon avalanche device at $V=-5$ V with optically excited carriers seeded at energy $E_{ph}$ in the conduction band on the $p$-side of the device. The local density of states in Fig.~\ref{fig:device_ldos_occ}(a) identifies the available electronic states throughout the channel. The bright spectral features trace the conduction- and valence-band structure, while their spatial tilt reflects the built-in and applied electrostatic field across the junction. The contact electrochemical potentials $\mu_p$ and $\mu_n$ define the nonequilibrium injection window imposed by the $p$- and $n$-type leads.

The occupied density of states in Fig.~\ref{fig:device_ldos_occ}(b), computed from the electron correlation function $G^<$, shows which of the available LDOS states are actually populated by contact injection. Unlike the LDOS, this quantity contains information about carrier occupation under bias. The occupied spectral weight follows the field-tilted band structure but is restricted by the contact distributions and the nonequilibrium boundary conditions. The photon-assisted reference energy $E_v+E_{ph}$ indicates the energy scale relevant for optically generated or promoted carriers in the avalanche process.

Together, these panels connect the open-system NEGF calculation to the physical requirements for impact ionization. Avalanche multiplication requires more than the presence of available states: a high-energy carrier must be populated, an occupied partner state must be available for scattering, and final states must exist after energy transfer. The LDOS establishes the allowed spectral structure of the biased device, while the occupied DOS identifies the subset of that structure participating in nonequilibrium transport. These results therefore show that the tight-binding Hamiltonian, lead self-energies, electrostatic band bending, and correlation functions combine to produce a consistent microscopic description of the silicon avalanche structure.

\subsection{Discussion}
The results presented above establish the microscopic transport baseline needed for modeling avalanche multiplication in a silicon SPAD-like structure. Instead of starting from an empirical ionization coefficient, the NEGF workflow constructs the device response from the fitted silicon Hamiltonian, open-boundary lead self-energies, electrostatic band bending, and nonequilibrium carrier occupations. The tight-binding band structure provides a compact but physically meaningful representation of the low-energy silicon states that dominate injection and transport near the band gap. Coupling to semi-infinite contacts then broadens these states, converting the isolated band structure into an open quantum system in which carriers can be injected, accelerated, scattered, and collected.

The position- and energy-resolved LDOS and occupied-DOS maps at $V=-5$ V give the most direct interpretation of the biased avalanche device. The LDOS identifies where electronic states are available in the channel and shows the field-induced tilt of the conduction- and valence-band features across the depletion region. The occupied DOS, computed from $G^<$, indicates the presence of avalanche buildup due to the incorporation of the the impact ionization terms $\Sigma_{II}^{<,>}$.


For the broader project, these results show that the numerical framework can combine atomistic electronic structure, contact coupling, applied bias, and NEGF correlation functions into a consistent open-device description. This is the required input for evaluating the impact-ionization self-energy $\Sigma_{II}$ self-consistently, so that carrier multiplication can eventually be predicted from the simulated spectral distribution rather than imposed through a local field-dependent rate. In the context of quantum networking, this matters because detector performance directly affects measurement fidelity, timing precision, dark counts, and detection probability in QKD, entanglement swapping, teleportation, and photonic computing protocols. A microscopic model of avalanche onset can therefore help connect silicon detector design choices to system-level quantum-network performance.

The present study is still limited in several important ways. First, we note that while the compressed $s$+$p$ hamiltonian used in this study was necessary for computational efficiency, it had a limited capacity to model high conduction band energies and indirect band gaps in silicon, which are necessary for more accurately modeling avalanche physics. Also, the device electrostatic profile was imposed rather than obtained from a fully self-consistent electron orbital-based Poisson--NEGF solution, and the device geometry used was a simplified p--i--n-like structure rather than a complete SPAD layout with guard rings, interfaces, defects, and realistic optical absorption.
Future work will therefore focus on incorporating larger Hamiltonian bases, incorporating Poisson electrostatics, adding phonon and defect scattering, and benchmarking the model against experimental avalanche statistics. These extensions will allow the framework to move from a transport baseline toward predictive design rules for integrated silicon single-photon detectors in quantum-network receivers.

\section{Conclusion}
This work addressed the need for a microscopic, quantum-transport description of avalanche multiplication in silicon single-photon detector structures, where conventional semiclassical models can obscure the energy-resolved and nonequilibrium processes that govern carrier multiplication. Using an atomistic NEGF framework, we constructed a silicon avalanche device model that combines a fitted tight-binding Hamiltonian, open-boundary contact self-energies, electrostatic band bending, and correlation functions that distinguish available states from occupied states. The resulting spectra show how the biased junction forms a field-tilted electronic structure and how contact injection populates selected regions of that structure, providing the physical phase space required for impact ionization.

These results establish the transport foundation for a self-consistent impact-ionization model in which avalanche gain can be computed from the nonequilibrium electronic structure rather than imposed through empirical ionization coefficients. For quantum networking, this is significant because detector behavior directly affects measurement fidelity, timing precision, dark counts, and the probability that a single absorbed photon becomes a usable electrical event. By linking atomistic device physics to the microscopic conditions for carrier multiplication, the NEGF approach developed here provides a pathway toward predictive modeling of silicon SPADs and APDs for integrated quantum receivers. The central outcome is that avalanche detection can be treated not only as a device-level gain process, but as a quantum-transport problem whose spectral structure, occupations, and scattering pathways can be resolved from first principles.

\subsection*{Data Availability Statement}

The data that support the ﬁndings of this study are openly available in the following github repository:\\
\href{https://github.com/cburdine/si-apd-modeling}{https://github.com/cburdine/si-apd-modeling}.

\subsection*{Acknowledgments}

The authors would like to thank Mike Hutcheson in the Baylor High-Performance Research and Computing Services department and Dr. Jake Minich in the Baylor University Department of Biology for granting dedicated access to GPU hardware used for the simulations conducted in this study. The authors also acknowledge the use of ChatGPT Codex for basic code review, code documentation, and help with unit testing.


\appendix

\label{app:negf_review}
\subsection{Review of Non-Equilibrium Green's Function Theory}

In this appendix section, we provide a brief review of general NEGF theory for quantum many-body systems and the energy-resolved Green's functions used in the main text. We begin by considering an interacting electronic system described by the second-quantized Hamiltonian of the form shown in \eqnref{eqn:Hamiltonian} in the main text. The Heisenberg equations of motion of the fermionic creation and annihilation operators $\hat{c}_i, \hat{c}_i^{\dagger}$ are given by
\begin{equation}
  \begin{aligned}
    i\hbar \frac{d}{dt}\hat{c}_n(t)
    &= [\hat{c}_n(t),\hat{H}_0(t)]
    + [\hat{c}_n(t),\hat{H}_{int}(t)] \\
    &= \sum_m h_{nm}\hat{c}_m(t) + \hat{F}_n(t).
    \label{eqn:c_heisenberg}
  \end{aligned}
\end{equation}
In \eqnref{eqn:c_heisenberg}, we introduce the interaction-induced ``force'' operators $\hat{F}_n$, which can be intuitively written as
\begin{equation}
  \hat{F}_n(t) =
  [\hat{c}_n(t),\hat{H}_{int}(t)]
  =
  \frac{\delta \hat{H}_{int}}{\delta \hat{c}_n(t)} .
\end{equation}

Solving these operator differential equations is quite difficult due to the complexity of the many-body interactions captured in the $\hat{F}_n(t)$ operators. The key idea of NEGF theory is to solve for Green's functions that propagate the correlations of the second-quantized field and force operators forward in time.

\subsubsection{Classical Green's functions}

Before introducing many-body Green's functions, it is helpful to first review classical Green's functions for independent, non-interacting fields and then generalize to the many-body case. For a classical, non-interacting system with a vector of independent field modes $c(t)$, we aim to solve the differential equation
\begin{equation}
  (i\partial_t - H)c(t) = F(t)
\end{equation}
where $H$ and $F(t)$ are matrices expanded in the basis of the field modes $c(t)$. Using classical Green's function methods, this equation is found to have the solution
\begin{equation}
  c(t) = c^0(t) + \int dt' G(t,t')F(t'),
  \label{eqn:classical_gf_solution}
\end{equation}
where $c^0(t) = \exp(-it H/\hbar)$ is the free field evolution and $G(t,t')$ is a Green's function that satisfies the defining equation
\begin{equation}
  (i\partial_t - H)G(t,t') = \delta(t-t').
  \label{eqn:classical_gf_delta}
\end{equation}
As \eqnref{eqn:classical_gf_delta} suggests, $G(t,t')$
serves as an ``inverse'' of the differential operator $(i\partial_t - H)$. Therefore, applying $G$ to the externally imposed force $F(t)$ recovers the solution in \eqnref{eqn:classical_gf_solution}.

\subsubsection{Quantum Many-Body Green's functions}

Unlike the classical case, the operators $\hat{F}_n(t)$ do not occupy the single-particle subspace, so we must resort to applying a mean-field approximation to make the dynamics tractable. We do this by restricting the equations of motion to the expectation values of observable two-time correlation functions:

\begin{equation}
  {\footnotesize
    \begin{aligned}
      i\hbar \dfrac{\partial}{\partial t}\expval{\hat{c}_a^{\dagger}(t') \hat{c}_n(t)} = \sum_m h_{nm} \expval{\hat{c}_a^{\dagger}(t')\hat{c}_m(t)} + \expval{\hat{c}_a^{\dagger}(t') \hat{F}_n(t)}
  \end{aligned}}
  \normalsize
  \tag{\normalsize\theequation}
  \label{eqn:correlation_eom}
\end{equation}

we introduce the matrices $G^{<}$ and $G^{>}$, called the \textit{lesser} and \textit{greater Green's functions}:
\begin{align}
  G_{ab}^{<}(t,t') &= i\expval{\hat{c}_b^{\dagger}(t')\hat{c}_a(t)} \\
  G_{ab}^{>}(t,t') &= -i\expval{\hat{c}_a(t)\hat{c}_b^{\dagger}(t')}
\end{align}
The lesser function describes \textit{occupied} electron states, while the greater function describes \textit{unoccupied} states (i.e., holes).

Likewise, we introduce the analogous matrices of correlation functions for the force operators, called the \textit{lesser} and \textit{greater self-energies}:
\begin{align}
  \Sigma_{ab}^{<}(t,t') &=
  \frac{i}{\hbar}\langle \hat{F}_b^\dagger(t')\hat{c}_a(t)\rangle, \\
  \Sigma_{ab}^{>}(t,t') &=
  \frac{-i}{\hbar}\langle \hat{c}_a(t)\hat{F}_b^\dagger(t')\rangle .
\end{align}
These quantities describe how interactions inject particles into states and remove particles from states.

Using these matrices of correlation operators along with the single-particle hamiltonian matrix $H$ as a basis for solving for the dynamics of a second-quantized system, we re-write \eqnref{eqn:correlation_eom} in the compact form
\begin{equation}
  \left( i\dfrac{\partial }{\partial t} - H\right)G^{<} = \Sigma^{R}G^< + \Sigma^< G^A
\end{equation}

The Green's functions that solve this equation yield both retarded solutions ($G^R$) and advanced solutions ($G^A$), which exhibit strict causality and anti-causality, respectively. These solutions assume the form
\begin{align}
  G_{ab}^R(t,t') &= -i\Theta(t-t')\langle \{\hat{c}_a(t),\hat{c}_b^\dagger(t')\} \rangle \\
  G_{ab}^A(t,t') &= i\Theta(t'-t)\langle \{\hat{c}_a(t),\hat{c}_b^\dagger(t')\} \rangle
\end{align}
where $\Theta(t)$ is the Heaviside step function, and $\{ A, B\} = AB + BA$ is the anticommutator. Combined, these functions satisfy the Green's function relation
\begin{align}
  &\left(i\dfrac{\partial}{\partial t} - H\right)G^{R,A}(t,t') = \delta(t-t') + \Big[ \\
  &\qquad \int dt_1\, \Sigma^{R,A}(t,t_1) G^{R,A}(t_1,t') \Big].
\end{align}

In the case when the system hamiltonian $\hat{H}$, is time-independent, the Green's functions $G^{<,>}(t,t')$ and $G^{R,A}(t,t')$ are time-invariant, and thus only depend on the time difference (e.g., $G(t,t') \equiv G(t-t')$). For these systems, it is advantageous to Fourier transform to the energy domain. The retarded Green's function then assumes the much simpler form shown in \eqnref{eqn:dyson}. Additionally, in the energy domain, the advanced and retarded solutions are Hermitian adjoints. For this reason, we drop the $R$/$A$ superscript in favor of using $G$/$G^\dagger$ and $\Sigma$/$\Sigma^\dagger$ to denote advanced/retarded solutions:
\begin{align}
  G(E) &:= G^{R}(E) = (G^{A}(E))^\dagger \\
  \Sigma(E) &:= \Sigma^{R}(E) = (\Sigma^{A}(E))^\dagger
\end{align}

In this representation, the equation of motion for $G^{<,>}$ simplifies quite dramatically, and one obtains the Keldysh relations (Equations \eqref{eqn:keldysh1} and \eqref{eqn:keldysh2}).

\bibliographystyle{IEEEtran}
\bibliography{references}


\end{document}